\documentclass[floatfix,%
 reprint,
 amsmath,amssymb,
 aps,
pra,
]{revtex4-2}

\usepackage{graphicx}
\usepackage{dcolumn}
\usepackage{bm}
\usepackage{physics}
\usepackage{placeins}
\usepackage{booktabs}
\usepackage{array}
\usepackage{subcaption}
\usepackage{enumitem} 
\usepackage{booktabs} 
\usepackage{siunitx}
\usepackage{subcaption}
\usepackage{tabularx} 
\usepackage{multirow} 

\begin{document}

\newcommand{\Caforty}{$^{40}$Ca } 
\newcommand{\Cafortytwo}{$^{42}$Ca } 
\newcommand{\Cafortythree}{$^{43}$Ca } 
\newcommand{\Cafortyfour}{$^{44}$Ca } 
\newcommand{\Cafortyeight}{$^{48}$Ca } 
\newcommand{\Cafortyion}{$^{40}\mathrm{Ca}^{+}$ } 
\newcommand{\Cafortythreeion}{$^{43}\mathrm{Ca}^{+}$ } 
\newcommand{\red}[1]{\textcolor{red}{#1}}

\preprint{APS/123-QED}

\title{A Comparison of Calcium Sources for Ion-Trap Loading via Laser Ablation}
\author{Daisy R H Smith} \author{Silpa Muralidharan} \author{Roland Hablützel}\altaffiliation[Current affiliation: ]{Nu Quantum Ltd., Cambridge, United Kingdom}
\author{Georgina Croft} \author{Klara Theophilo} \author{Alexander Owens} \author{Yashna N D Lekhai} \author{Scott J Thomas} \author{Cameron Deans}%
 \email{Corresponding address: cameron.deans@stfc.ac.uk}
\affiliation{National Quantum Computing Centre, Didcot OX11 0QX, United Kingdom}


\date{\today}

\begin{abstract}
Trapped-ion technology is a leading approach for scalable quantum computing. A key element of ion trapping is reliable loading of atomic sources into the trap. While thermal atomic ovens have traditionally been used for this purpose, laser ablation has emerged as a viable alternative in recent years, offering the advantages of faster and more localized loading with lower heat dissipation. Calcium is a well-established ion for qubit applications. Here we examine a range of calcium sources for ablation and provide a comprehensive analysis of each. We consider factors such as ease of use, temperature and yield of the ablation plume, and the lifetime of ablation spots. For each target, we estimate the number of trappable atoms per ablation pulse for a typical surface and 3D ion trap.
\end{abstract}

\maketitle


\section{Introduction}

Quantum computing is an emerging technology with the potential to transform industries such as pharmaceuticals and manufacturing \cite{cao2018potential, nielsen2010quantum}. Trapped-ion quantum computing is a promising approach, with state of the art single and two-qubit fidelities, high above the error correction threshold \cite{srinivas2021high, gaebler2016high, leu2023fast, weber2024robust, loschnauer_2024}. Trapped ions offer advantages as qubits due to their atomic uniformity and all-to-all connectivity. These properties have led to numerous proposals for scalable architectures \cite{kielpinski2002architecture, moses2023race, lekitsch2017blueprint, loschnauer_2024, mordini2024multi}.

Several different ions can be used for quantum computing, including \Cafortyion and $^{43}\text{Ca}^{+}$, which exhibit high fidelity in both single- and two-qubit gates across various schemes \cite{bruzewicz2019trapped, ballance2016high, benhelm2008towards, schafer2018fast, clark2021high}, as well as high state-preparation and measurement fidelity \cite{harty_2014}. Mixed-species interactions and sympathetic cooling have been successfully demonstrated with $^{9}\text{Be}^{+}$ and $^{88}\text{Sr}^{+}$ \cite{lancellotti2023low, hughes2020benchmarking} and the required lasers for Doppler cooling are readily available solid-state lasers, not requiring frequency doubling.

Ion traps have traditionally been loaded using atomic ovens, which resistively heat sources to release atoms into the trapping zone \cite{kjaergaard2000isotope, david_lucas_2004}. This approach has several drawbacks, including slow ion-loading, coating of the ion-trap electrodes, generation of magnetic fields, and high thermal load applied to the trap \cite{leibrandt_2007, hendricks2007all}. Another approach is ablation, where a pulsed laser creates a small atom plume from a target. Ablation offers several advantages, including a more localized source, faster and more precise timing of ion loading, and reduced thermal load applied to the ion trap \cite{sheridan2011all, vrijsen_2019, leibrandt_2007, hendricks2007all, brendan_white_2022, battles_2024}.

A further advantage of ablation loading is the ability to use compound targets, which are often easier and cheaper to procure than pure elemental sources. Targets vary in key characteristics such as surface reflectivity and density which can affect the performance of ablation \cite{sheridan2011all}. Previously, a range of targets have been compared for $^{88}\text{Sr}^{+}$ \cite{leibrandt_2007}, $^{138}\text{Ba}^{+}$ \cite{shi2023ablation}, and pure calcium has been compared with $\text{CaTiO}_{3}$ \cite{battles_2024}. 

Here, we present a study of six potential calcium targets and compare them by ease of preparation and mounting in the vacuum chamber and analysis of the plume in terms of yield and temperature. The yield and temperature values are calculated via time-resolved spectroscopy experiments on the $\text{S}_{0}$ $\rightarrow$ $\text{P}_{1}$ first-stage photoionization transition, and used to estimate the number of trappable atoms in the plume for each target for a representative surface and 3D trap. We also present digital microscope images of the targets after ablation.

\section{Target Selection and Preparation}

\begin{table*}[ht]
\caption{Comparison of ablation targets and their preparation.}
\label{tab:target}
\centering
\begin{tabular}{ p{3.8cm} p{2.8cm} p{2.1cm} p{1.8cm} p{0.6cm} p{5.8cm} }
\toprule
\textbf{Target} & \textbf{Structure} & \textbf{Adhesive} & \raggedleft \textbf{\% Calcium} &\ & \textbf{Comments}  \\
\midrule
Black calcite & Bulk crystal & Epoxy & \raggedleft 33.6\% &\ &  Easy to mount\\
White calcite & Powder & Indium & \raggedleft 39.6\% &\ & Easy to mount\\
Pure calcium & Bulk metal & Indium & \raggedleft 99.0\% &\ & Oxidizes in ambient conditions\\
Calcium titanate & Powder & Indium & \raggedleft 29.5\% &\ & Difficult to mount\\
Calcium carbide & Bulk crystal & Epoxy & \raggedleft 50.0\% &\ & Disintegrates after thermal cycle\\
White calcite & Bulk crystal & N/A & \raggedleft 13.7\% &\ & Does not glue with epoxy and indium \\
\bottomrule
\end{tabular}
\end{table*}

Considerations when selecting the targets for comparison included the materials' availability, robustness, stability under UHV and cryogenic conditions, as well as ease of mounting. Some materials are prone to fracture or ignition at low temperatures, so careful consideration was given to avoid potential disturbances to pressure or temperature, which could affect the trapping of ions or subsequent experiments. Both bulk crystal and powdered targets were compared in terms of mounting and ablation performance.

Selected targets were mounted to the system using one of two adhesives: a vacuum-compatible epoxy (EPO-TEK H20E) or pure indium foil (thickness 0.1 mm, $\geq$ 99.995\% trace metals basis), which was melted and applied as glue. The targets are fabricated with approximate dimensions of 10 mm x 5 mm, with varying thicknesses for each individual target. Considering the beam size and required spacing, this allows for around 350 ablation spots per target. 

Table \ref{tab:target} provides a comparison of the various targets, detailing their mounting methods and preparation specifics, used to assess their suitability. 

\textbf{Black calcite (Si:CaCO$_3$)}: Silicon-doped calcite crystals were bonded to the ablation mount using epoxy. This target is easy to mount and stable at room temperature and under vacuum conditions.

\textbf{White calcite (CaCO$_3$)}: White calcite powder was bonded to the ablation mount using indium. A thin layer of the powder was formed on the mount. The material remains stable at room temperature and under vacuum conditions. Bulk white calcite failed to adhere during gluing due to the high temperatures involved in both epoxy and indium bonding, which caused the crystals to fracture.

\textbf{Pure calcium (Ca)}: Pure calcium metal was selected for its high potential yield but requires careful handling under vacuum as it oxidizes in ambient conditions. Such targets may therefore need to be replaced when breaking the vacuum \cite{battles_2024}. This motivates the use of chemically inert salts as alternatives and is particularly relevant when using expensive isotopically enriched targets. The target was mounted using indium, and initial handling was performed in a glove box to prevent oxidation. The target was stored under vacuum prior to use. 

\textbf{Calcium titanate (CaTiO$_3$)}: Calcium titanate did not bond effectively with the ablation mount, although a thin layer of the material was successfully applied for testing. The material remains stable at room temperature and under vacuum conditions.

\textbf{Calcium carbide (CaC$_2$)}: Calcium carbide, a bulk crystal, was bonded using epoxy. It required handling in a glove box to prevent oxidation or ignition during preparation. While stable at both room and cryogenic temperatures, heating and cooling caused the target to shatter, rendering it unsuitable for long-term use.

\section{Experimental Setup}

Ablation experiments were conducted in a vacuum chamber equipped with cryogenic cooling. The chamber is initially evacuated to a pressure of around $10^{-6}$ mbar and cooled to 3.3 K over a 24-hour period. Cryo-pumping results in a pressure below $10^{-11}$ mbar at the experimental region.

\begin{figure}[h!]
\includegraphics[width=0.8\columnwidth]{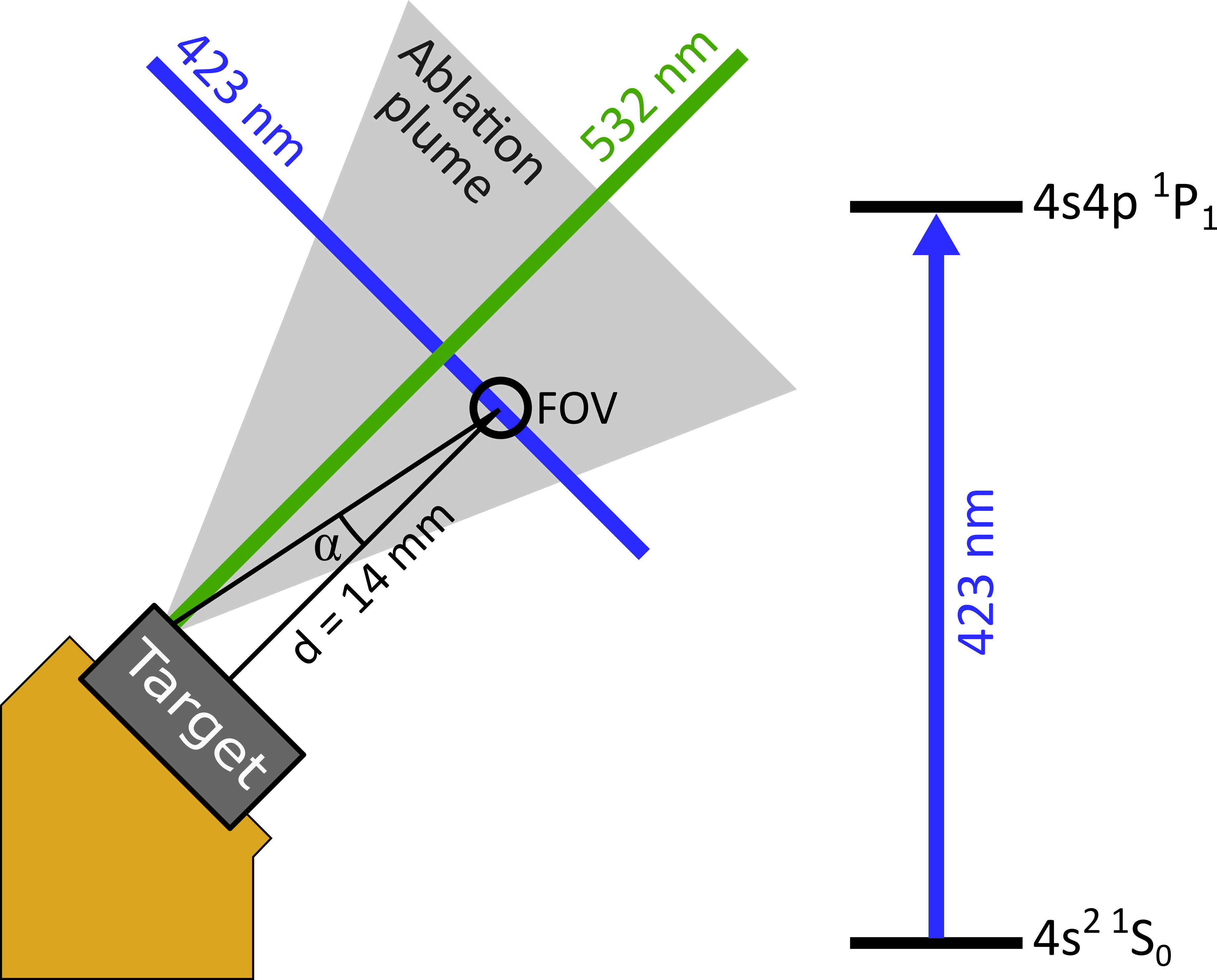}
\caption{\label{fig:beam_set_up} The experimental setup, showing the \SI{532}{\nano\meter} (ablation) beam, the \SI{423}{\nano\meter} (fluorescence) beam. FOV (field of view) refers to the area visible to the PMT. In reality this region is smaller than the diameter of the \SI{423}{\nano\meter} beam. The angle $\alpha$ is the angle that an atom in the ablation plume makes with the normal to the fluorescence beam and $d$ is the distance between the target surface and the FOV. The relevant calcium transition is shown on the right. }
\end{figure}

The target is mounted on a gold-plated copper removable mount that is subsequently mounted inside the experimental region of the vacuum chamber. This allowed for fast changing of different test materials and easy replacement of targets. Ablation is performed using a 532 nm Nd:YAG pulsed laser (Bright Solutions WEDGE HB), and fluorescence detection is performed with a 423 nm CW laser (Toptica MDL Pro HP). The laser setup is illustrated in Figure \ref{fig:beam_set_up}. Fluorescence from the ablation plume is detected by a photomultiplier tube (PMT). 

The 532 nm laser has a pulse duration of 1.1 ns and a maximum pulse energy of 1 mJ. It is focused onto the target using an external lens mounted on an XYZ translation stage, allowing precise movement and control over the ablation spot. The spot is approximately 400 $\mu$m. Band pass filters at 423 nm are placed before the PMT to eliminate unwanted scattering, ensuring that only the fluorescence at 423 nm is detected. The 423 nm laser beam, with a diameter of 1.8 mm, is continuously applied. The laser wavelength is tuned to the resonant wavelength of the $4\text{s}^{2}$ $^{1}\text{S}_{0}$ $\rightarrow$ $4\text{s}4\text{p}^{1} \text{P}_{1}$ transition in $^{40}$Ca ($\lambda_\text{vac} = 422.7920$ nm) \cite{NIST_ASD}. The power is chosen such that the intensity is significantly higher than 10 times the saturation intensity. The fluence of the 532 nm laser is gradually increased until the first ablation signal is detected. The fluence at which fluorescence is first detected is defined as the ablation threshold. Each target showed a different ablation threshold, as shown in Table \ref{tab:target_results}.

\section{Data analysis}\label{sec:data_analysis}

\begin{figure*}[ht]
    \centering
    \includegraphics[width=\textwidth]{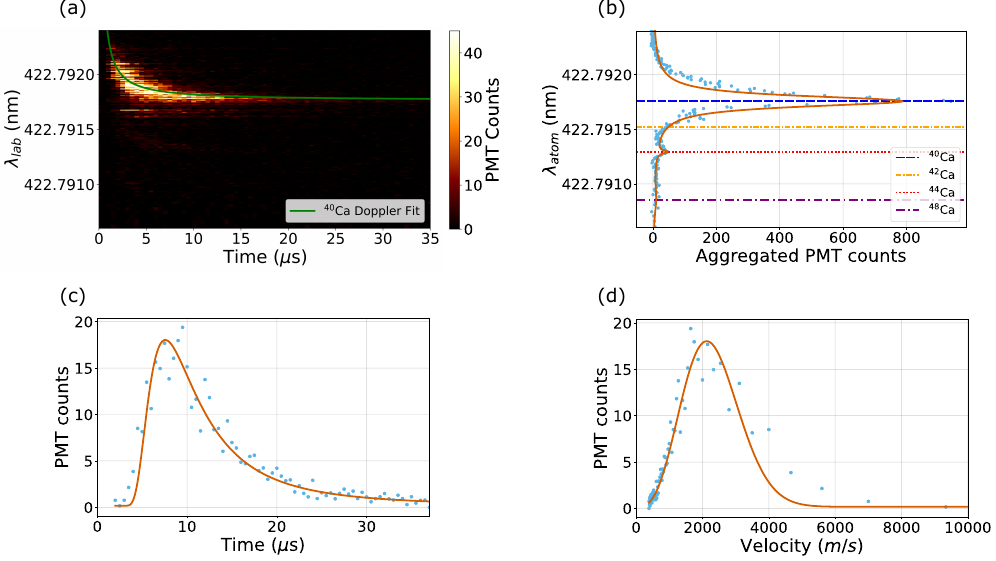}
    \caption{An example to show how the data for a ablation experiment was analyzed to obtain temperature and yield. This data was taken with the calcium carbide target. (a) Fluorescence counts scanned over time and wavelength ($\lambda_{lab}$). The data is fitted with Equation \ref{eqn:doppler_shift_hyperbola} for angle ($\alpha$) and resonant wavelength ($\lambda_{atom}$) of the photoionization transition for $^{40}$Ca. (b) The sum of PMT counts along the hyperbola (Equation~\ref{eqn:doppler_shift_hyperbola}) for a given wavelength, fitted with Equation \ref{eqn:isotope_lorentzian}. (c) A time-of-flight distribution generated by the fluorescence along the hyperbola in (a), fitted for temperature with a Maxwell-Boltzmann distribution curve (Equation \ref{eqn:MB_TOF}). (d) The atom velocity distribution, derived from Figure (c), included for illustrative purposes.}
    \label{fig:four_part}
\end{figure*}

Time-resolved spectroscopy was used to compare the temperature and yield of the ablation plume across the different targets. The experimental procedure involves setting a fixed wavelength of the \SI{423}{\nano\meter} laser, triggering a pulse from the 532 nm laser, and measuring the fluorescence from the \SI{423}{\nano\meter} photoionization transition over a \SI{35}{\micro\second} interval using a PMT with a binning of \SI{0.5}{\micro\second}. The process is repeated across a range of wavelengths, forming a plot such as that seen in Figure \ref{fig:four_part}(a).

The data is used to calculate the angle that the ablation plume makes with the normal to the fluorescence beam, via fitting the equation

\begin{equation}\label{eqn:doppler_shift_hyperbola}
\lambda_{lab}(\lambda_{atom}, \alpha, t) = \lambda_{atom}
\left(1 - \frac{d \sin(\alpha)}{c \, t }\right) ,
\end{equation}%
 where $\lambda_{lab}$ is the wavelength of the incident \SI{423}{\nano\meter} laser, $\lambda_{atom}$ is the wavelength of the \Caforty $\text{S}_{0}$ $\rightarrow$ $\text{P}_{1}$ transition in the atom frame, $\alpha$ is the angle between the atom's trajectory and the normal to the fluorescence beam as seen in Figure \ref{fig:beam_set_up}, $t$ is the time since the ablation laser pulse, $d$ is the distance from the center of the field of view and the surface of the ablation target (\SI{14}{\milli\meter}), and $c$ is the speed of light in a vacuum. \Caforty is assumed to produce the highest fluorescence, as it has the highest natural abundance of 96.9\% \cite{david_lucas_2004}. Fitting Equation \ref{eqn:doppler_shift_hyperbola} to the brightest pixel per wavelength, determines $\alpha$ and $\lambda_{atom}$ for \Caforty, as seen in Figure \ref{fig:four_part}(a). 

Given $\alpha$ (which is fixed for all isotopes and a given spot), one can construct from Equation \ref{eqn:doppler_shift_hyperbola}, the hyperbola of data point which correspond to a given $\lambda_{atom}$. We sum along these hyperbola to produce Figure \ref{fig:four_part}(b). This data is fitted with a sum of Lorentzians, one for each of the four most abundant calcium isotopes \cite{david_lucas_2004} 

\begin{equation}\label{eqn:isotope_lorentzian}
L_{\text{total}}(\lambda) = \sum_{i=0}^{3} K_i \cdot \frac{\frac{s_i \cdot \gamma_i}{2}}{1 + s_i + \left( \frac{2\delta}{\gamma_i} \right)^2} \, ,
\end{equation}%
where $\delta$ is the detuning from the resonant frequency of the $\text{S}_{0}$ $\rightarrow$ $\text{P}_{1}$ transition, \( s_i \), \( \gamma_i \), and \( K_i \) are respectively the saturation parameter, linewidth, and scaling factor for isotopes \Caforty, \Cafortytwo, \Cafortyfour, and \Cafortyeight. \Caforty consistently appears as the most abundant isotope, with \Cafortyfour, which has the second-highest abundance (2.09\%), typically observed as a minor peak. The other isotopes, which have abundances below 1\% \cite{david_lucas_2004}, are not clearly detected.

To determine the temperature of the ablation plume, we plot the time values and corresponding fluorescence along the \Caforty hyperbola (Equation~\ref{eqn:doppler_shift_hyperbola}) from Figure \ref{fig:four_part}(a), generating the time-of-flight data shown in \ref{fig:four_part}(c). This is then fitted with a Maxwell-Boltzmann distribution. As the processed data is generated by accounting for the Doppler shift, which only takes into account atom movement in one direction, a one-dimensional Maxwell-Boltzmann distribution,

\begin{equation}
\text{MB}(v) = A \exp( \frac{- m \, v^2}{2 \, k_B T })\, ,
\end{equation}%
is used. The corresponding equation for time-of-flight is derived (see Appendix \ref{app:derivation}) as

\begin{equation}\label{eqn:MB_TOF}
\text{MB}(t) = -A \frac{d}{t^3} \exp\left( \frac{- m \, d^2}{2 \, k_B t^2 T} \right) ,
\end{equation}%
and is used to fit Figure \ref{fig:four_part}(c) to obtain a temperature value. The geometry of the system and dimensions of the targets are such that $\alpha$ is small ($\arccos{(\alpha)}\approx 1$). Therefore, we fix $d=$\SI{14}{\milli\meter} throughout to simplfy the analysis. The corresponding Maxwell-Boltzmann distribution over velocity is also shown for illustrative purposes in Figure \ref{fig:four_part}(d). 

\section{Results}\label{sec:results}

The data analysis process as described in the previous section was conducted for all ablation spots on each target. Consistent with other literature \cite{brendan_white_2022}, the number of successful spots (identified by the successful fitting of a clear fluorescence curve for $^{40}$Ca) varied across targets. The spots that produced clear fluorescence curves were collated to obtain the following yield and temperature results.

Yield and plume temperature are key metrics for comparing ablation targets. A high yield indicates a dense population of calcium atoms, increasing the likelihood of successful trapping. A low plume temperature corresponds to a higher population of slower atoms, which are easier to trap. This is particularly relevant in surface traps, which typically have a lower trap depth than 3D traps \cite{auchter_2022, weber_2023}.

Yield was determined by summing the total fluorescence and subtracting the background fluorescence in the time-resolved spectroscopy plots. The background fluorescence was determined by averaging the photon counts from the top-right quadrant of the time-resolved spectroscopy plot, an area that consistently showed no atomic signal.

\begin{figure}[h]
\centering
\includegraphics[width=\columnwidth]{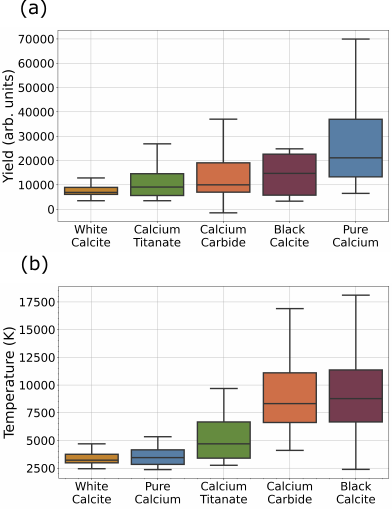}
\caption{(a) Box-and-whisker plot presenting the spread of ablation plume yield for each target. (b) Plot presenting the spread of ablation plume temperature for each target.} 
\label{fig:yield_temp}
\end{figure}

The results for yield are shown in Figure \ref{fig:yield_temp}(a). The box-and-whisker plot shows that pure calcium exhibits the highest yield of 21000, while the two powders, white calcite and calcium titanate, show the lowest yields, with medians of 6900, and 9100 respectively. This is in-keeping with what we expect given metallic calcium is the most calcium-rich target and the bulk samples have lower surface roughness. 

\begin{table*}
\caption{Comparison of ablation targets and results. Threshold refers to the fluence at which the ablation pulse first showed fluorescence. The temperatures and yield values are the median temperature and yield for each target as seen in Figures \ref{fig:yield_temp}(a) and \ref{fig:yield_temp}(b).}
\label{tab:target_results}
\setlength{\tabcolsep}{10pt} 
\renewcommand{\arraystretch}{1.2} 
\begin{tabular}{l r r r}
\toprule
\textbf{Target} & \textbf{Threshold (mJ/cm$^2$)} & \textbf{Temperature (K)} & \textbf{Yield (arb. units)} \\ 
\midrule
White Calcite & 0.028 & 3200 & 6900 \\
Pure Calcium &  0.013 & 3500 & 21000 \\
Calcium Titanate &  0.028 & 4700 & 9100 \\
Calcium Carbide & 0.059 & 8300 & 10000 \\
Black Calcite  & 0.108 & 8800 & 15000 \\
\bottomrule
\end{tabular}
\end{table*}

The temperature of the ablation plume was calculated using a Maxwell-Boltzmann distribution fit, as discussed in Section \ref{sec:data_analysis}. A box-and-whisker plot displays the results for each of the targets in Figure \ref{fig:yield_temp}(b). It can be seen that white calcite and pure calcium produced the plumes with the lowest temperatures, with medians of 3200 K, and 3500 K. The temperatures for pure calcium and calcium titanate are similar to previously reported values \cite{battles_2024}.

\begin{figure}[h!]
\includegraphics[width=0.9\columnwidth]{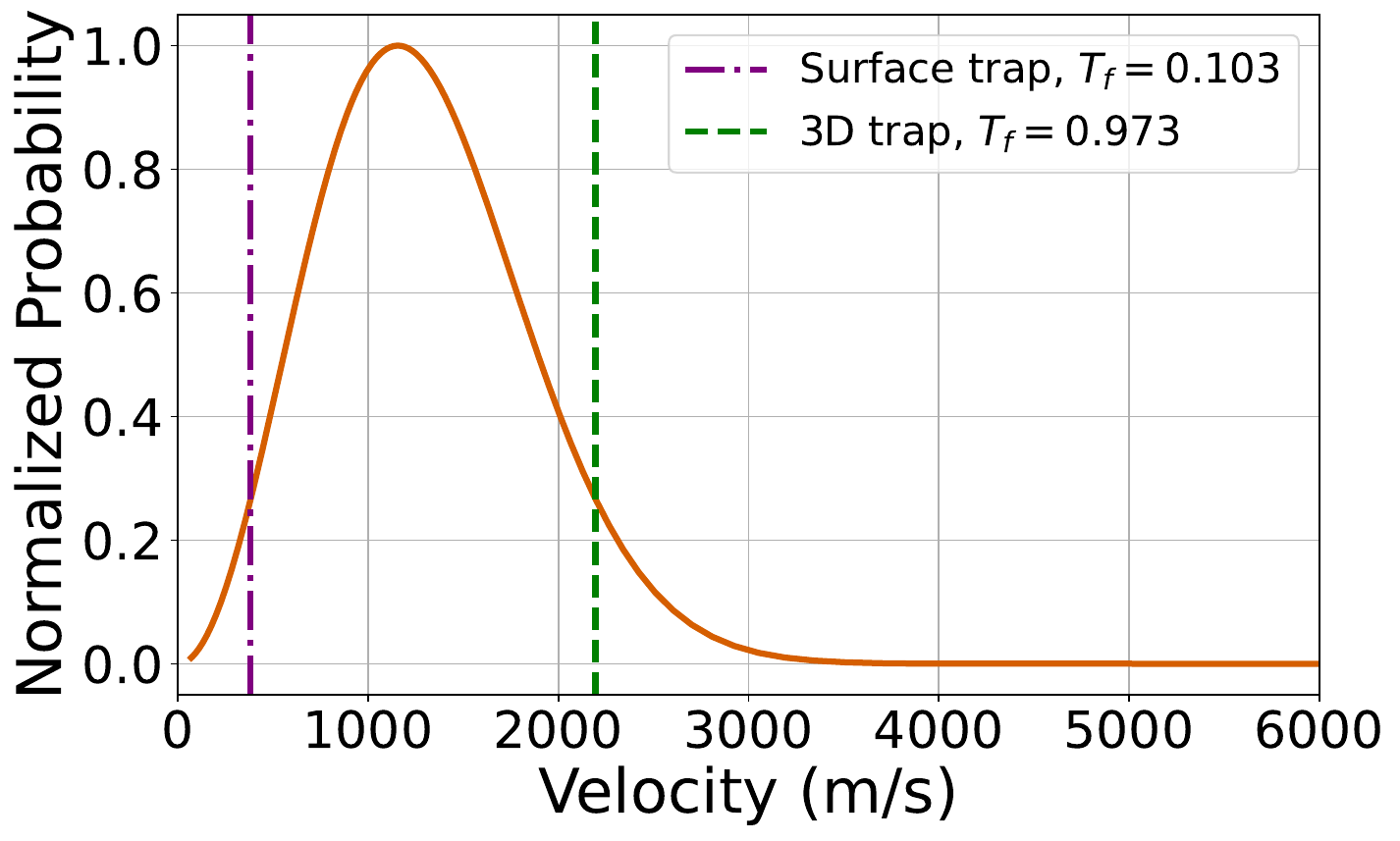}
\caption{\label{fig:trappable_atoms_white_calcite} Extrapolated Maxwell-Boltzmann distribution curve for white calcite, used to determine the fraction of trappable atoms in the ablation plume using the trap-depth for an example 3D trap and surface trap. $T_{f}$ is the trappable fraction of the atoms.}
\end{figure}

Using the calculated median temperature for each target, we estimate the fraction of trappable atoms for a surface and 3D ion trap. Example trap-depths of \SI{30}{\milli\electronvolt} and \SI{1}{\electronvolt} were chosen for a surface and 3D trap, respectively \cite{weber_2023, auchter_2022}. The maximum velocity of a trappable atom was calculated from the trap depth, and the fraction of atoms below that velocity was established using the temperature (see Appendix \ref{app:trappable_atoms_for_all_targets}). The results are shown in Figure \ref{fig:trappable_atoms_white_calcite} for white calcite, which had the lowest temperature and therefore the highest fraction of trappable atoms. The results for all targets are displayed in Table \ref{tab:trappable_atoms}.

The fraction of trappable atoms was used to estimate a total number of trappable atoms for each target as described in Appendix \ref{app:trappable_atoms_for_all_targets}. The results are displayed in Table \ref{tab:trappable_atoms}. For surface traps, pure calcium and white calcite had the highest estimated trappable atoms, while for 3D traps, the best targets were pure calcium and black calcite. This reflects the greater influence of temperature in surface traps, whereas in 3D traps, because the trap depth is higher, the yield of the target is of higher importance.  

\begin{table*}
\caption{Estimated fraction of trappable atoms in a typical ablation plume for a typical surface and 3D ion trap, determined using the median plume temperature for each target. The total number of trappable atoms is calculated from fluorescence-based estimates of the plume's atomic population (see Appendix \ref{app:trappable_atoms_for_all_targets}). }
\label{tab:trappable_atoms}
\setlength{\tabcolsep}{10pt} 
\renewcommand{\arraystretch}{1.2} 
\begin{tabular}{l cc|cc} 
\toprule
\textbf{} & \multicolumn{2}{c}{\textbf{Fraction Trappable}} & \multicolumn{2}{c}{\textbf{Estimated Total Trappable Atoms}} \\ 
\cmidrule(lr){2-3} \cmidrule(lr){4-5}
\textbf{Target} & \textbf{Surface Trap} & \textbf{3D Trap} & \textbf{Surface Trap} & \textbf{3D Trap} \\ 
\midrule
White Calcite & 0.103 & 0.97 & 42 & 400 \\
Pure Calcium &  0.096 & 0.97 & 120 & 1200 \\
Calcium Titanate &  0.072 & 0.92 & 38 & 490 \\
Calcium Carbide & 0.041 & 0.75 & 24 & 440 \\
Black Calcite  & 0.039 & 0.73 & 34 & 650 \\
\bottomrule
\end{tabular}
\end{table*}

\begin{figure*}
\includegraphics[width=\textwidth]{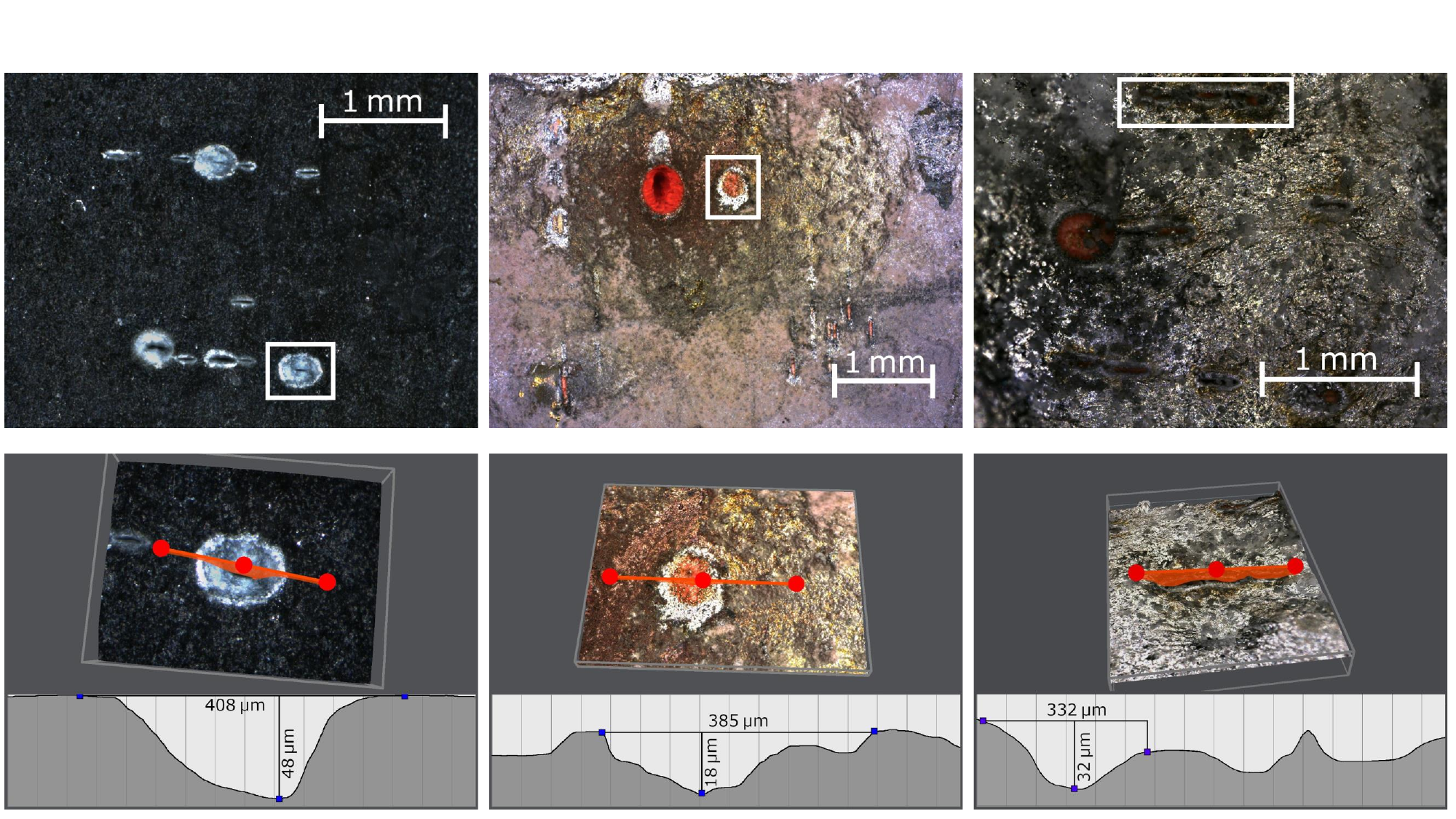}
\caption{\label{fig:metrology} Images taken using a digital microscope to examine target-surfaces after ablation. Left panels: black calcite (bulk crystal). Middle panels: calcium titanate (powder). Right panels: white calcite (powder). }
\end{figure*}

Another parameter of interest is spot lifetime, defined as the number of ablation pulses a target spot can endure before it becomes unusable, requiring realignment of the ablation beam to a new location. The primary factor influencing this parameter is whether the target is a powder or a bulk crystal. This likely reflects the preparation of the target and not its chemical properties. For the powders the lifetime was established by determining the point at which indium (the mount-target bonding agent) became visible. This was evident through an increase in the fluorescence across the entire time-resolved spectroscopy plot, which made the Doppler-shift hyperbola harder to detect (see Appendix \ref{app:indium_plot}). This is attributed to strong resonance lines of singly-ionized indium at at \SI{423}{\nano\meter} \cite{lang1931erste}. The lifetime of white calcite and calcium titanate were thus seen to be, on average, 1900 and 850 pulses respectively. 

The lifetime of the bulk crystals were not directly observed. However, 4500, 6300, and 3000 pulses were applied for calcium carbide, black calcite, and pure calcium, respectively, without exhausting the ablation spot. 

It should also be noted that the powder-spots were more inconsistent. For white calcite and calcium titanate there was a respective 31.8\% and 30.4\% probability of seeing no fluorescence at all from a spot, whereas for the bulk crystals 100\% of the spots showed fluorescence. We attribute this to the thinner and more uneven surface of the powdered targets, as seen in Figure \ref{fig:metrology}, reducing the likelihood that a given spot consists of a sufficiently thick layer of the calcium target for reliable successful ablation.

A digital microscope (Evocam II) was used to examine the targets after ablation. The width and depth of the ablation spots were measured for black calcite, white calcite and calcium titanate. Post-ablation imaging of the calcium carbide and pure calcium targets was not possible, as the targets crumbled and oxidized, respectively, upon removal from the cryogenic vacuum system. Examples of the images obtained are shown in Figure \ref{fig:metrology}. The bulk crystal (black calcite) displays clear circular craters, whereas the powders show a more uneven surface around the ablation spots. This is attributed to the less rigid structure of the pasted powders as compared to the solid material. Figure \ref{fig:metrology} shows that the crater widths and depths are of the same order of magnitude across targets. The widths are attributed to the beam size, and therefore appear similar between targets. The red color of the ablation spots on the powder targets is due to the underlying copper oxide becoming exposed.

\section{Conclusion}

Ablation loading of ion traps has several advantages relevant to a range quantum technologies applications. These include reduced thermal loads, localized atom sources, and fast precisely timed loading. We examined six potential ablation targets for calcium ion-trap loading. Black calcite and white calcite were found to be the most practical for UHV mounting. Ablation was performed with black calcite, white calcite powder, pure calcium, calcium titanate, and calcium carbide. Examination of the spot-lifetime showed that powders had shorter lifetimes than bulk crystals. Digital microscope images and analysis of bulk crystal and powder targets highlight the factors contributing to this discrepancy. Time-resolved spectroscopy was used to analyze the yield and temperature of the ablation plumes. Black calcite and pure calcium produced the highest yield, while white calcite and pure calcium had the lowest temperature. This was reflected in the estimation of the number of trappable atoms per ablation plume, where pure calcium and white calcite were preferable surface traps, while pure calcium and black calcite were preferable for 3D traps. However, the preparation environment and exposure duration of a particular experiment may eliminate pure calcium due to oxidation. The target may have to be cleaned with a series of high-power pulses to remove the oxide layer following each system modification \cite{shi2023ablation} or replaced entirely \cite{battles_2024}. In this case, we conclude that white calcite powder would be the most suitable target material for surface traps and black calcite would be the most suitable for 3D traps. This extends to other species of fast-oxidizing metals, such as strontium and barium, and is of particular importance when considering expensive isotopically enriched targets. 

\begin{acknowledgments}

The authors acknowledge the support of Connor Pettitt from the NQCC Software and Control Systems team, Harry Byrne, Charles Evans, Sam Allum, and Dave Wilsher from the STFC Technology Department, and James Taylor from STFC ISIS Neutron and Muon Source. We are also grateful to Esme Somerside Gregory for helpful discussions on the data analysis. This work was supported by the UK National Quantum Computing Centre [NQCC200921]. 

\end{acknowledgments}

\appendix

\section{Derivation of the Maxwell-Boltzmann time-of-flight fitting equation}\label{app:derivation}

We consider only one dimension because the processed data accounts for the Doppler shift which only applies to the direction of the laser propagation. The standard Maxwell-Boltzmann distribution equation for velocity in one dimension is
\begin{equation}
\text{MB}(v) = A \exp( \frac{- m \, v^2}{2 \, k_B T })\, ,
\end{equation}
were $m$ is the mass of the atom, $T$ is the temperature of the ablation plume, $k_B$ is the Boltzmann constant, and $A$ is a scaling factor. We use the chain rule to rearrange the equation in terms of time-of-flight
\begin{equation}
\text{MB}(t) = \text{MB}(t) \times \frac{dv}{dt} \, ,
\end{equation}
\begin{equation}
\frac{dv}{dt} = \frac{d}{dt} \left[\frac{d}{t}\right] = -\frac{d}{t^2} \, ,
\end{equation}
where $d$ represents the distance between the target and the PMT's field of view, and $t$ denotes the time-of-flight, the duration taken by the atoms to travel from the target to the field of view. To account for the fact that slower atoms spend more time in the field of view and therefore contribute more fluorescence, we introduce a factor of $\frac{1}{t}$ and get 
\begin{equation}
\text{MB}(t) = -A \frac{d}{t^3} \exp\left( \frac{- m \, d^2}{2 \, k_B t^2 T} \right) ,
\end{equation}%
which is used to fit the time-of-flight data for a temperature value in Section \ref{sec:data_analysis}. 
\nocite{*}

\section{Example of a time-resolved spectroscopy plot with indium fluorescence}\label{app:indium_plot}

When a powder spot was depleted, fluorescence from indium dominated the signal. This is attributed to strong resonant lines of In II at \SI{423}{\nano\meter} \cite{lang1931erste}. Upon removing the targets from the vacuum for inspection, the ablation spot was found to have eroded down to the indium layer, but not the copper beneath, confirming indium as the fluorescence source.
\begin{figure}[h]
\includegraphics[width=0.9\columnwidth]{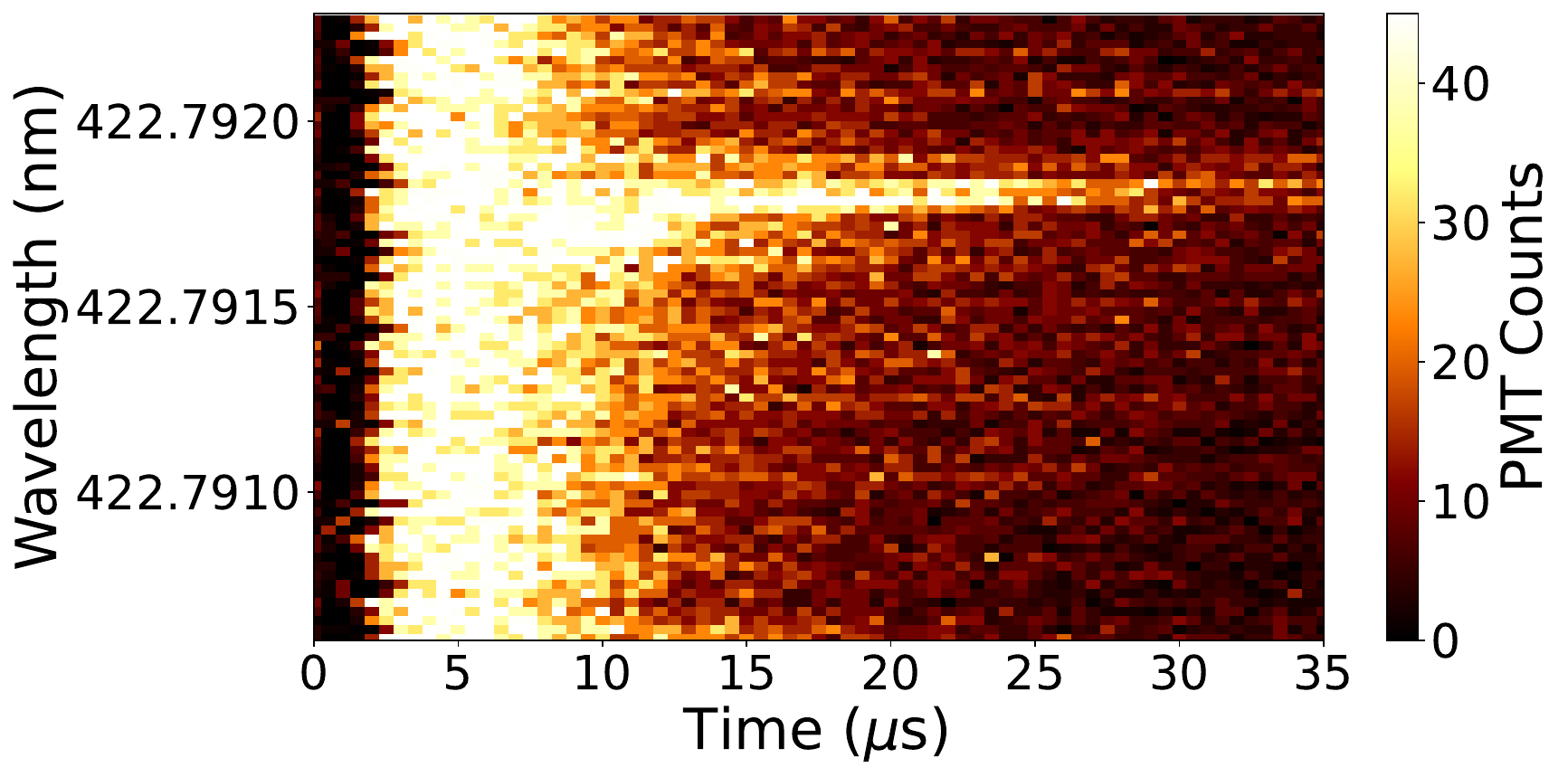}
\caption{Example time-resolved spectroscopy plot where the ablation spot as been eroded down to the indium bonding material. The fluorescence counts are too high to distinguish the Doppler-shift hyperbola accurately. This is attributed to the presence of resonant lines of singly-ionized indium.}\label{fig:example_fig_2D_calcium_titanate_indium}
\end{figure}

\section{Trappable atoms for all targets}\label{app:trappable_atoms_for_all_targets}
\begin{figure*}[b]
\includegraphics[width=\textwidth]{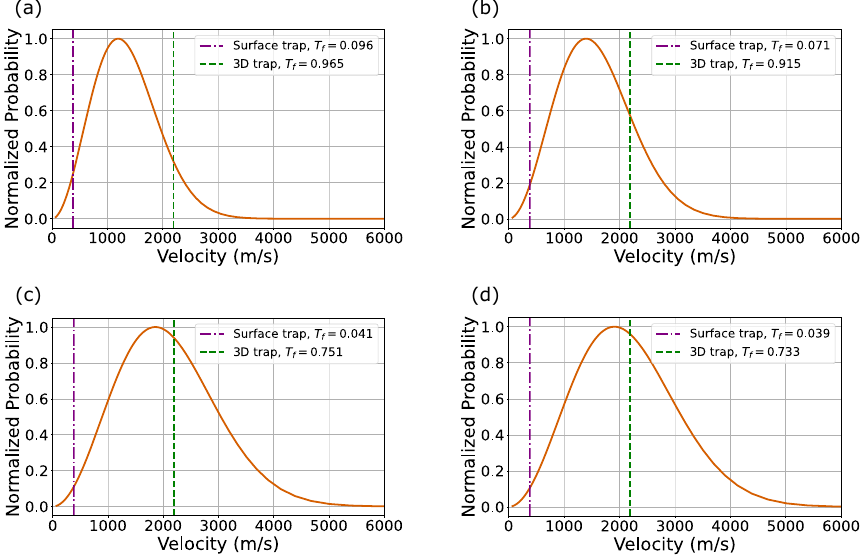}
\caption{\label{fig:four_part_trappable_atoms_plot} Maxwell-Boltzmann distributions for each target, generated from the median temperatures. Used to determine the fraction of trappable atoms in the ablation plume using the trap-depths for a typical 3D and surface trap. $T_{f}$ is the trappable fraction of atoms. (a) Pure calcium. (b) Calcium titanate. (c) Calcium carbide. (d) Black calcite.}
\end{figure*}

The maximum velocity of a trappable atom is calculated using equation
\begin{equation}
v_{\text{max}} = \sqrt{\frac{2E_{\text{trap}}}{m}} \, , 
\end{equation}
where $E_{\text{trap}}$ is the trap depth and $m$ is the mass of a calcium atom. From $v_{\text{max}}$ the minimum time-of-flight for trappable atoms can be calculated, and thus the fraction of trappable atoms is determined via 
\begin{equation}
T_{f} = \frac{\int_{\text{TOF}_{\min}}^{\infty} \frac{d}{t^3} \exp\left(-\frac{m d^2}{2 t^2 k_B T}\right) dt}{\int_{0}^{\infty} \frac{d}{t^3} \exp\left(-\frac{m d^2}{2 t^2 k_B T}\right) dt} \, .
\end{equation}
The maximum velocity of a trappable atom is shown on a Maxwell-Boltzmann distribution for each target, generated using the median temperature calculated for each target, in Figures \ref{fig:trappable_atoms_white_calcite} and \ref{fig:four_part_trappable_atoms_plot}.

\begin{table*}[h]
    \caption{Number of trappable atoms from the ablation plume, for a single laser pulse, for each target type. Calculated using the trap depth of a typical surface ion trap and a typical 3D ion trap. Ranges calculated using the quartiles of the corresponding yield and temperature data.}
    \label{tab:trappable_atom_calc}
    \centering
    \begin{tabular}{l@{\quad\quad}c@{\quad\quad}c@{\quad\quad}c@{\quad\quad}}
        \toprule
        \multicolumn{4}{c}{\textbf{Surface Ion Trap}} \\
        \cmidrule(lr){1-4}
        \textbf{Target} & Fraction Trappable & Number of Atoms & Total Trappable \\
        \midrule
        White Calcite    & 0.103 &  406 &  42 [36, 45]    \\
        Pure Calcium     & 0.096 & 1234  & 118 [99, 143]   \\
        Calcium Titanate & 0.072 &  535  &  38 [27, 52]   \\
        Calcium Carbide  & 0.041 &  587  &  24 [18, 30]   \\
        Black Calcite    & 0.039 &  882  &  34 [27, 45]   \\
        \midrule
        \multicolumn{4}{c}{\textbf{3D Ion Trap}} \\
        \cmidrule(lr){1-4}
        \textbf{Target} & Fraction Trappable & Number of Atoms & Total Trappable \\
        \midrule
        White Calcite    & 0.973 &  406  &  395 [387, 397]  \\
        Pure Calcium     & 0.965 & 1234  & 1191 [1159, 1214]  \\
        Calcium Titanate & 0.915 &  535  &  489 [441, 517]  \\
        Calcium Carbide  & 0.751 &  587  &  442 [381, 486]  \\
        Black Calcite    & 0.733 &  882  &  646 [564, 727]  \\
        \bottomrule
    \end{tabular}
\end{table*}

Given the estimated collection efficiency of the imaging set-up ($5\%$, based on the solid-angle calculation and estimation of the beam path efficiency to the PMT) and the PMT's quantum efficiency (31\%), we estimate the overall photon-collection efficiency as $\frac{P_{PMT}}{P_A} = 0.02$, where $P_{PMT}$ is the number of photons collected by the PMT and $P_A$ is the total number of photons produced by the atom cloud within the FOV.

Using a representative atom velocity of \SI{2000}{\meter\per\second}, based on the peak velocities shown in Figure \ref{fig:four_part_trappable_atoms_plot}, and a fluorescence beam width of \SI{1.8}{\milli\meter}, we estimate the total fluorescence duration of an atom as $t = \frac{\SI{1.8}{\milli\meter}}{\SI{2000}{\meter\per\second}} = \SI{0.9}{\micro\second}$.

The scattering rate of the \SI{423}{\nano\meter} transition is approximately \SI{220}{\mega\hertz} \cite{NIST_ASD}. With an estimated excited-state population of \SI{40}{\percent}, each atom scatters photons at a rate of $s_{r} = 0.4 \times \SI{220}{\mega\hertz} = \SI{88}{\mega\hertz}$. Over the fluorescence duration, the total number of scattered photons is given by $t \times s_{r} = \SI{0.9}{\micro\second} \times \SI{88}{\mega\hertz} = 79$ photons per atom. This results in about $0.01$ atoms per photon at the atom cloud. We calculate the number of atoms per photon seen at the PMT to be
\begin{equation}
    \frac{A}{P_{PMT}} = \frac{A}{P_{A}} \times \frac{P_{A}}{P_{PMT}} = 0.01 \times \frac{1}{0.02} = 0.5 \, , 
\end{equation}
\\
where $A$ is the number of atoms in the atom cloud, $P_{PMT}$ is the total number of photons collected by the PMT, $P_A$ is the total number of photons produced by the atom cloud.

To estimate the total number of trappable atoms, the median yield for each target was used to scale a typical time-resolved spectroscopy plot so that its total fluorescence matched the median yield. The brightest row in the plot was then assumed to represent the total fluorescence from the ablation plume when illuminated near resonance, for a single ablation pulse. This was multiplied by $\frac{A}{P_{PMT}}$ to estimate the number of atoms in the ablation plume, and then by the trappable fraction ($T_{f}$) to estimate the total number of trappable atoms. Error ranges are calculated using the quartiles of the corresponding yield and temperature data. The results are seen in Table \ref{tab:trappable_atom_calc}. 


\bibliography{references}

\begin{thebibliography}{35}%
\makeatletter
\providecommand \@ifxundefined [1]{%
 \@ifx{#1\undefined}
}%
\providecommand \@ifnum [1]{%
 \ifnum #1\expandafter \@firstoftwo
 \else \expandafter \@secondoftwo
 \fi
}%
\providecommand \@ifx [1]{%
 \ifx #1\expandafter \@firstoftwo
 \else \expandafter \@secondoftwo
 \fi
}%
\providecommand \natexlab [1]{#1}%
\providecommand \enquote  [1]{``#1''}%
\providecommand \bibnamefont  [1]{#1}%
\providecommand \bibfnamefont [1]{#1}%
\providecommand \citenamefont [1]{#1}%
\providecommand \href@noop [0]{\@secondoftwo}%
\providecommand \href [0]{\begingroup \@sanitize@url \@href}%
\providecommand \@href[1]{\@@startlink{#1}\@@href}%
\providecommand \@@href[1]{\endgroup#1\@@endlink}%
\providecommand \@sanitize@url [0]{\catcode `\\12\catcode `\$12\catcode `\&12\catcode `\#12\catcode `\^12\catcode `\_12\catcode `\%12\relax}%
\providecommand \@@startlink[1]{}%
\providecommand \@@endlink[0]{}%
\providecommand \url  [0]{\begingroup\@sanitize@url \@url }%
\providecommand \@url [1]{\endgroup\@href {#1}{\urlprefix }}%
\providecommand \urlprefix  [0]{URL }%
\providecommand \Eprint [0]{\href }%
\providecommand \doibase [0]{https://doi.org/}%
\providecommand \selectlanguage [0]{\@gobble}%
\providecommand \bibinfo  [0]{\@secondoftwo}%
\providecommand \bibfield  [0]{\@secondoftwo}%
\providecommand \translation [1]{[#1]}%
\providecommand \BibitemOpen [0]{}%
\providecommand \bibitemStop [0]{}%
\providecommand \bibitemNoStop [0]{.\EOS\space}%
\providecommand \EOS [0]{\spacefactor3000\relax}%
\providecommand \BibitemShut  [1]{\csname bibitem#1\endcsname}%
\let\auto@bib@innerbib\@empty
\bibitem [{\citenamefont {Cao}\ \emph {et~al.}(2018)\citenamefont {Cao}, \citenamefont {Romero},\ and\ \citenamefont {Aspuru-Guzik}}]{cao2018potential}%
  \BibitemOpen
  \bibfield  {author} {\bibinfo {author} {\bibfnamefont {Y.}~\bibnamefont {Cao}}, \bibinfo {author} {\bibfnamefont {J.}~\bibnamefont {Romero}},\ and\ \bibinfo {author} {\bibfnamefont {A.}~\bibnamefont {Aspuru-Guzik}},\ }\href@noop {} {\bibfield  {journal} {\bibinfo  {journal} {IBM Journal of Research and Development}\ }\textbf {\bibinfo {volume} {62}},\ \bibinfo {pages} {6} (\bibinfo {year} {2018})}\BibitemShut {NoStop}%
\bibitem [{\citenamefont {Nielsen}\ and\ \citenamefont {Chuang}(2010)}]{nielsen2010quantum}%
  \BibitemOpen
  \bibfield  {author} {\bibinfo {author} {\bibfnamefont {M.~A.}\ \bibnamefont {Nielsen}}\ and\ \bibinfo {author} {\bibfnamefont {I.~L.}\ \bibnamefont {Chuang}},\ }\href@noop {} {\emph {\bibinfo {title} {Quantum computation and quantum information}}}\ (\bibinfo  {publisher} {Cambridge university press},\ \bibinfo {year} {2010})\BibitemShut {NoStop}%
\bibitem [{\citenamefont {Srinivas}\ \emph {et~al.}(2021)\citenamefont {Srinivas}, \citenamefont {Burd}, \citenamefont {Knaack}, \citenamefont {Sutherland}, \citenamefont {Kwiatkowski}, \citenamefont {Glancy}, \citenamefont {Knill}, \citenamefont {Wineland}, \citenamefont {Leibfried}, \citenamefont {Wilson} \emph {et~al.}}]{srinivas2021high}%
  \BibitemOpen
  \bibfield  {author} {\bibinfo {author} {\bibfnamefont {R.}~\bibnamefont {Srinivas}}, \bibinfo {author} {\bibfnamefont {S.}~\bibnamefont {Burd}}, \bibinfo {author} {\bibfnamefont {H.}~\bibnamefont {Knaack}}, \bibinfo {author} {\bibfnamefont {R.}~\bibnamefont {Sutherland}}, \bibinfo {author} {\bibfnamefont {A.}~\bibnamefont {Kwiatkowski}}, \bibinfo {author} {\bibfnamefont {S.}~\bibnamefont {Glancy}}, \bibinfo {author} {\bibfnamefont {E.}~\bibnamefont {Knill}}, \bibinfo {author} {\bibfnamefont {D.}~\bibnamefont {Wineland}}, \bibinfo {author} {\bibfnamefont {D.}~\bibnamefont {Leibfried}}, \bibinfo {author} {\bibfnamefont {A.~C.}\ \bibnamefont {Wilson}}, \emph {et~al.},\ }\href@noop {} {\bibfield  {journal} {\bibinfo  {journal} {Nature}\ }\textbf {\bibinfo {volume} {597}},\ \bibinfo {pages} {209} (\bibinfo {year} {2021})}\BibitemShut {NoStop}%
\bibitem [{\citenamefont {Gaebler}\ \emph {et~al.}(2016)\citenamefont {Gaebler}, \citenamefont {Tan}, \citenamefont {Lin}, \citenamefont {Wan}, \citenamefont {Bowler}, \citenamefont {Keith}, \citenamefont {Glancy}, \citenamefont {Coakley}, \citenamefont {Knill}, \citenamefont {Leibfried} \emph {et~al.}}]{gaebler2016high}%
  \BibitemOpen
  \bibfield  {author} {\bibinfo {author} {\bibfnamefont {J.~P.}\ \bibnamefont {Gaebler}}, \bibinfo {author} {\bibfnamefont {T.~R.}\ \bibnamefont {Tan}}, \bibinfo {author} {\bibfnamefont {Y.}~\bibnamefont {Lin}}, \bibinfo {author} {\bibfnamefont {Y.}~\bibnamefont {Wan}}, \bibinfo {author} {\bibfnamefont {R.}~\bibnamefont {Bowler}}, \bibinfo {author} {\bibfnamefont {A.~C.}\ \bibnamefont {Keith}}, \bibinfo {author} {\bibfnamefont {S.}~\bibnamefont {Glancy}}, \bibinfo {author} {\bibfnamefont {K.}~\bibnamefont {Coakley}}, \bibinfo {author} {\bibfnamefont {E.}~\bibnamefont {Knill}}, \bibinfo {author} {\bibfnamefont {D.}~\bibnamefont {Leibfried}}, \emph {et~al.},\ }\href@noop {} {\bibfield  {journal} {\bibinfo  {journal} {Physical review letters}\ }\textbf {\bibinfo {volume} {117}},\ \bibinfo {pages} {060505} (\bibinfo {year} {2016})}\BibitemShut {NoStop}%
\bibitem [{\citenamefont {Leu}\ \emph {et~al.}(2023)\citenamefont {Leu}, \citenamefont {Gely}, \citenamefont {Weber}, \citenamefont {Smith}, \citenamefont {Nadlinger},\ and\ \citenamefont {Lucas}}]{leu2023fast}%
  \BibitemOpen
  \bibfield  {author} {\bibinfo {author} {\bibfnamefont {A.}~\bibnamefont {Leu}}, \bibinfo {author} {\bibfnamefont {M.}~\bibnamefont {Gely}}, \bibinfo {author} {\bibfnamefont {M.}~\bibnamefont {Weber}}, \bibinfo {author} {\bibfnamefont {M.}~\bibnamefont {Smith}}, \bibinfo {author} {\bibfnamefont {D.}~\bibnamefont {Nadlinger}},\ and\ \bibinfo {author} {\bibfnamefont {D.}~\bibnamefont {Lucas}},\ }\href@noop {} {\bibfield  {journal} {\bibinfo  {journal} {Physical Review Letters}\ }\textbf {\bibinfo {volume} {131}},\ \bibinfo {pages} {120601} (\bibinfo {year} {2023})}\BibitemShut {NoStop}%
\bibitem [{\citenamefont {Weber}\ \emph {et~al.}(2024)\citenamefont {Weber}, \citenamefont {Gely}, \citenamefont {Hanley}, \citenamefont {Harty}, \citenamefont {Leu}, \citenamefont {L{\"o}schnauer}, \citenamefont {Nadlinger},\ and\ \citenamefont {Lucas}}]{weber2024robust}%
  \BibitemOpen
  \bibfield  {author} {\bibinfo {author} {\bibfnamefont {M.}~\bibnamefont {Weber}}, \bibinfo {author} {\bibfnamefont {M.}~\bibnamefont {Gely}}, \bibinfo {author} {\bibfnamefont {R.}~\bibnamefont {Hanley}}, \bibinfo {author} {\bibfnamefont {T.}~\bibnamefont {Harty}}, \bibinfo {author} {\bibfnamefont {A.}~\bibnamefont {Leu}}, \bibinfo {author} {\bibfnamefont {C.}~\bibnamefont {L{\"o}schnauer}}, \bibinfo {author} {\bibfnamefont {D.}~\bibnamefont {Nadlinger}},\ and\ \bibinfo {author} {\bibfnamefont {D.}~\bibnamefont {Lucas}},\ }\href@noop {} {\bibfield  {journal} {\bibinfo  {journal} {Physical Review A}\ }\textbf {\bibinfo {volume} {110}},\ \bibinfo {pages} {L010601} (\bibinfo {year} {2024})}\BibitemShut {NoStop}%
\bibitem [{\citenamefont {L{\"o}schnauer}\ \emph {et~al.}(2024)\citenamefont {L{\"o}schnauer}, \citenamefont {Toba}, \citenamefont {Hughes}, \citenamefont {King}, \citenamefont {Weber}, \citenamefont {Srinivas}, \citenamefont {Matt}, \citenamefont {Nourshargh}, \citenamefont {Allcock}, \citenamefont {Ballance} \emph {et~al.}}]{loschnauer_2024}%
  \BibitemOpen
  \bibfield  {author} {\bibinfo {author} {\bibfnamefont {C.}~\bibnamefont {L{\"o}schnauer}}, \bibinfo {author} {\bibfnamefont {J.~M.}\ \bibnamefont {Toba}}, \bibinfo {author} {\bibfnamefont {A.}~\bibnamefont {Hughes}}, \bibinfo {author} {\bibfnamefont {S.}~\bibnamefont {King}}, \bibinfo {author} {\bibfnamefont {M.}~\bibnamefont {Weber}}, \bibinfo {author} {\bibfnamefont {R.}~\bibnamefont {Srinivas}}, \bibinfo {author} {\bibfnamefont {R.}~\bibnamefont {Matt}}, \bibinfo {author} {\bibfnamefont {R.}~\bibnamefont {Nourshargh}}, \bibinfo {author} {\bibfnamefont {D.}~\bibnamefont {Allcock}}, \bibinfo {author} {\bibfnamefont {C.}~\bibnamefont {Ballance}}, \emph {et~al.},\ }\href@noop {} {\bibfield  {journal} {\bibinfo  {journal} {arXiv preprint arXiv:2407.07694}\ } (\bibinfo {year} {2024})}\BibitemShut {NoStop}%
\bibitem [{\citenamefont {Kielpinski}\ \emph {et~al.}(2002)\citenamefont {Kielpinski}, \citenamefont {Monroe},\ and\ \citenamefont {Wineland}}]{kielpinski2002architecture}%
  \BibitemOpen
  \bibfield  {author} {\bibinfo {author} {\bibfnamefont {D.}~\bibnamefont {Kielpinski}}, \bibinfo {author} {\bibfnamefont {C.}~\bibnamefont {Monroe}},\ and\ \bibinfo {author} {\bibfnamefont {D.~J.}\ \bibnamefont {Wineland}},\ }\href@noop {} {\bibfield  {journal} {\bibinfo  {journal} {Nature}\ }\textbf {\bibinfo {volume} {417}},\ \bibinfo {pages} {709} (\bibinfo {year} {2002})}\BibitemShut {NoStop}%
\bibitem [{\citenamefont {Moses}\ \emph {et~al.}(2023)\citenamefont {Moses}, \citenamefont {Baldwin}, \citenamefont {Allman}, \citenamefont {Ancona}, \citenamefont {Ascarrunz}, \citenamefont {Barnes}, \citenamefont {Bartolotta}, \citenamefont {Bjork}, \citenamefont {Blanchard}, \citenamefont {Bohn} \emph {et~al.}}]{moses2023race}%
  \BibitemOpen
  \bibfield  {author} {\bibinfo {author} {\bibfnamefont {S.~A.}\ \bibnamefont {Moses}}, \bibinfo {author} {\bibfnamefont {C.~H.}\ \bibnamefont {Baldwin}}, \bibinfo {author} {\bibfnamefont {M.~S.}\ \bibnamefont {Allman}}, \bibinfo {author} {\bibfnamefont {R.}~\bibnamefont {Ancona}}, \bibinfo {author} {\bibfnamefont {L.}~\bibnamefont {Ascarrunz}}, \bibinfo {author} {\bibfnamefont {C.}~\bibnamefont {Barnes}}, \bibinfo {author} {\bibfnamefont {J.}~\bibnamefont {Bartolotta}}, \bibinfo {author} {\bibfnamefont {B.}~\bibnamefont {Bjork}}, \bibinfo {author} {\bibfnamefont {P.}~\bibnamefont {Blanchard}}, \bibinfo {author} {\bibfnamefont {M.}~\bibnamefont {Bohn}}, \emph {et~al.},\ }\href@noop {} {\bibfield  {journal} {\bibinfo  {journal} {Physical Review X}\ }\textbf {\bibinfo {volume} {13}},\ \bibinfo {pages} {041052} (\bibinfo {year} {2023})}\BibitemShut {NoStop}%
\bibitem [{\citenamefont {Lekitsch}\ \emph {et~al.}(2017)\citenamefont {Lekitsch}, \citenamefont {Weidt}, \citenamefont {Fowler}, \citenamefont {M{\o}lmer}, \citenamefont {Devitt}, \citenamefont {Wunderlich},\ and\ \citenamefont {Hensinger}}]{lekitsch2017blueprint}%
  \BibitemOpen
  \bibfield  {author} {\bibinfo {author} {\bibfnamefont {B.}~\bibnamefont {Lekitsch}}, \bibinfo {author} {\bibfnamefont {S.}~\bibnamefont {Weidt}}, \bibinfo {author} {\bibfnamefont {A.~G.}\ \bibnamefont {Fowler}}, \bibinfo {author} {\bibfnamefont {K.}~\bibnamefont {M{\o}lmer}}, \bibinfo {author} {\bibfnamefont {S.~J.}\ \bibnamefont {Devitt}}, \bibinfo {author} {\bibfnamefont {C.}~\bibnamefont {Wunderlich}},\ and\ \bibinfo {author} {\bibfnamefont {W.~K.}\ \bibnamefont {Hensinger}},\ }\href@noop {} {\bibfield  {journal} {\bibinfo  {journal} {Science Advances}\ }\textbf {\bibinfo {volume} {3}},\ \bibinfo {pages} {e1601540} (\bibinfo {year} {2017})}\BibitemShut {NoStop}%
\bibitem [{\citenamefont {Mordini}\ \emph {et~al.}(2024)\citenamefont {Mordini}, \citenamefont {Vasquez}, \citenamefont {Motohashi}, \citenamefont {M{\"u}ller}, \citenamefont {Malinowski}, \citenamefont {Zhang}, \citenamefont {Mehta}, \citenamefont {Kienzler},\ and\ \citenamefont {Home}}]{mordini2024multi}%
  \BibitemOpen
  \bibfield  {author} {\bibinfo {author} {\bibfnamefont {C.}~\bibnamefont {Mordini}}, \bibinfo {author} {\bibfnamefont {A.~R.}\ \bibnamefont {Vasquez}}, \bibinfo {author} {\bibfnamefont {Y.}~\bibnamefont {Motohashi}}, \bibinfo {author} {\bibfnamefont {M.}~\bibnamefont {M{\"u}ller}}, \bibinfo {author} {\bibfnamefont {M.}~\bibnamefont {Malinowski}}, \bibinfo {author} {\bibfnamefont {C.}~\bibnamefont {Zhang}}, \bibinfo {author} {\bibfnamefont {K.~K.}\ \bibnamefont {Mehta}}, \bibinfo {author} {\bibfnamefont {D.}~\bibnamefont {Kienzler}},\ and\ \bibinfo {author} {\bibfnamefont {J.~P.}\ \bibnamefont {Home}},\ }\href@noop {} {\bibfield  {journal} {\bibinfo  {journal} {arXiv preprint arXiv:2401.18056}\ } (\bibinfo {year} {2024})}\BibitemShut {NoStop}%
\bibitem [{\citenamefont {Bruzewicz}\ \emph {et~al.}(2019)\citenamefont {Bruzewicz}, \citenamefont {Chiaverini}, \citenamefont {McConnell},\ and\ \citenamefont {Sage}}]{bruzewicz2019trapped}%
  \BibitemOpen
  \bibfield  {author} {\bibinfo {author} {\bibfnamefont {C.~D.}\ \bibnamefont {Bruzewicz}}, \bibinfo {author} {\bibfnamefont {J.}~\bibnamefont {Chiaverini}}, \bibinfo {author} {\bibfnamefont {R.}~\bibnamefont {McConnell}},\ and\ \bibinfo {author} {\bibfnamefont {J.~M.}\ \bibnamefont {Sage}},\ }\href@noop {} {\bibfield  {journal} {\bibinfo  {journal} {Applied Physics Reviews}\ }\textbf {\bibinfo {volume} {6}} (\bibinfo {year} {2019})}\BibitemShut {NoStop}%
\bibitem [{\citenamefont {Ballance}\ \emph {et~al.}(2016)\citenamefont {Ballance}, \citenamefont {Harty}, \citenamefont {Linke}, \citenamefont {Sepiol},\ and\ \citenamefont {Lucas}}]{ballance2016high}%
  \BibitemOpen
  \bibfield  {author} {\bibinfo {author} {\bibfnamefont {C.~J.}\ \bibnamefont {Ballance}}, \bibinfo {author} {\bibfnamefont {T.~P.}\ \bibnamefont {Harty}}, \bibinfo {author} {\bibfnamefont {N.~M.}\ \bibnamefont {Linke}}, \bibinfo {author} {\bibfnamefont {M.~A.}\ \bibnamefont {Sepiol}},\ and\ \bibinfo {author} {\bibfnamefont {D.~M.}\ \bibnamefont {Lucas}},\ }\href@noop {} {\bibfield  {journal} {\bibinfo  {journal} {Physical review letters}\ }\textbf {\bibinfo {volume} {117}},\ \bibinfo {pages} {060504} (\bibinfo {year} {2016})}\BibitemShut {NoStop}%
\bibitem [{\citenamefont {Benhelm}\ \emph {et~al.}(2008)\citenamefont {Benhelm}, \citenamefont {Kirchmair}, \citenamefont {Roos},\ and\ \citenamefont {Blatt}}]{benhelm2008towards}%
  \BibitemOpen
  \bibfield  {author} {\bibinfo {author} {\bibfnamefont {J.}~\bibnamefont {Benhelm}}, \bibinfo {author} {\bibfnamefont {G.}~\bibnamefont {Kirchmair}}, \bibinfo {author} {\bibfnamefont {C.~F.}\ \bibnamefont {Roos}},\ and\ \bibinfo {author} {\bibfnamefont {R.}~\bibnamefont {Blatt}},\ }\href@noop {} {\bibfield  {journal} {\bibinfo  {journal} {Nature Physics}\ }\textbf {\bibinfo {volume} {4}},\ \bibinfo {pages} {463} (\bibinfo {year} {2008})}\BibitemShut {NoStop}%
\bibitem [{\citenamefont {Sch{\"a}fer}\ \emph {et~al.}(2018)\citenamefont {Sch{\"a}fer}, \citenamefont {Ballance}, \citenamefont {Thirumalai}, \citenamefont {Stephenson}, \citenamefont {Ballance}, \citenamefont {Steane},\ and\ \citenamefont {Lucas}}]{schafer2018fast}%
  \BibitemOpen
  \bibfield  {author} {\bibinfo {author} {\bibfnamefont {V.}~\bibnamefont {Sch{\"a}fer}}, \bibinfo {author} {\bibfnamefont {C.}~\bibnamefont {Ballance}}, \bibinfo {author} {\bibfnamefont {K.}~\bibnamefont {Thirumalai}}, \bibinfo {author} {\bibfnamefont {L.}~\bibnamefont {Stephenson}}, \bibinfo {author} {\bibfnamefont {T.}~\bibnamefont {Ballance}}, \bibinfo {author} {\bibfnamefont {A.}~\bibnamefont {Steane}},\ and\ \bibinfo {author} {\bibfnamefont {D.}~\bibnamefont {Lucas}},\ }\href@noop {} {\bibfield  {journal} {\bibinfo  {journal} {Nature}\ }\textbf {\bibinfo {volume} {555}},\ \bibinfo {pages} {75} (\bibinfo {year} {2018})}\BibitemShut {NoStop}%
\bibitem [{\citenamefont {Clark}\ \emph {et~al.}(2021)\citenamefont {Clark}, \citenamefont {Tinkey}, \citenamefont {Sawyer}, \citenamefont {Meier}, \citenamefont {Burkhardt}, \citenamefont {Seck}, \citenamefont {Shappert}, \citenamefont {Guise}, \citenamefont {Volin}, \citenamefont {Fallek} \emph {et~al.}}]{clark2021high}%
  \BibitemOpen
  \bibfield  {author} {\bibinfo {author} {\bibfnamefont {C.~R.}\ \bibnamefont {Clark}}, \bibinfo {author} {\bibfnamefont {H.~N.}\ \bibnamefont {Tinkey}}, \bibinfo {author} {\bibfnamefont {B.~C.}\ \bibnamefont {Sawyer}}, \bibinfo {author} {\bibfnamefont {A.~M.}\ \bibnamefont {Meier}}, \bibinfo {author} {\bibfnamefont {K.~A.}\ \bibnamefont {Burkhardt}}, \bibinfo {author} {\bibfnamefont {C.~M.}\ \bibnamefont {Seck}}, \bibinfo {author} {\bibfnamefont {C.~M.}\ \bibnamefont {Shappert}}, \bibinfo {author} {\bibfnamefont {N.~D.}\ \bibnamefont {Guise}}, \bibinfo {author} {\bibfnamefont {C.~E.}\ \bibnamefont {Volin}}, \bibinfo {author} {\bibfnamefont {S.~D.}\ \bibnamefont {Fallek}}, \emph {et~al.},\ }\href@noop {} {\bibfield  {journal} {\bibinfo  {journal} {Physical Review Letters}\ }\textbf {\bibinfo {volume} {127}},\ \bibinfo {pages} {130505} (\bibinfo {year} {2021})}\BibitemShut {NoStop}%
\bibitem [{\citenamefont {Harty}\ \emph {et~al.}(2014)\citenamefont {Harty}, \citenamefont {Allcock}, \citenamefont {Ballance}, \citenamefont {Guidoni}, \citenamefont {Janacek}, \citenamefont {Linke}, \citenamefont {Stacey},\ and\ \citenamefont {Lucas}}]{harty_2014}%
  \BibitemOpen
  \bibfield  {author} {\bibinfo {author} {\bibfnamefont {T.~P.}\ \bibnamefont {Harty}}, \bibinfo {author} {\bibfnamefont {D.~T.~C.}\ \bibnamefont {Allcock}}, \bibinfo {author} {\bibfnamefont {C.~J.}\ \bibnamefont {Ballance}}, \bibinfo {author} {\bibfnamefont {L.}~\bibnamefont {Guidoni}}, \bibinfo {author} {\bibfnamefont {H.~A.}\ \bibnamefont {Janacek}}, \bibinfo {author} {\bibfnamefont {N.~M.}\ \bibnamefont {Linke}}, \bibinfo {author} {\bibfnamefont {D.~N.}\ \bibnamefont {Stacey}},\ and\ \bibinfo {author} {\bibfnamefont {D.~M.}\ \bibnamefont {Lucas}},\ }\href {https://doi.org/10.1103/PhysRevLett.113.220501} {\bibfield  {journal} {\bibinfo  {journal} {Phys. Rev. Lett.}\ }\textbf {\bibinfo {volume} {113}},\ \bibinfo {pages} {220501} (\bibinfo {year} {2014})}\BibitemShut {NoStop}%
\bibitem [{\citenamefont {Lancellotti}\ \emph {et~al.}(2024)\citenamefont {Lancellotti}, \citenamefont {Welte}, \citenamefont {Simoni}, \citenamefont {Mordini}, \citenamefont {Behrle}, \citenamefont {de~Neeve}, \citenamefont {Marinelli}, \citenamefont {Negnevitsky},\ and\ \citenamefont {Home}}]{lancellotti2023low}%
  \BibitemOpen
  \bibfield  {author} {\bibinfo {author} {\bibfnamefont {F.}~\bibnamefont {Lancellotti}}, \bibinfo {author} {\bibfnamefont {S.}~\bibnamefont {Welte}}, \bibinfo {author} {\bibfnamefont {M.}~\bibnamefont {Simoni}}, \bibinfo {author} {\bibfnamefont {C.}~\bibnamefont {Mordini}}, \bibinfo {author} {\bibfnamefont {T.}~\bibnamefont {Behrle}}, \bibinfo {author} {\bibfnamefont {B.}~\bibnamefont {de~Neeve}}, \bibinfo {author} {\bibfnamefont {M.}~\bibnamefont {Marinelli}}, \bibinfo {author} {\bibfnamefont {V.}~\bibnamefont {Negnevitsky}},\ and\ \bibinfo {author} {\bibfnamefont {J.~P.}\ \bibnamefont {Home}},\ }\href {https://doi.org/10.1103/PhysRevResearch.6.L032059} {\bibfield  {journal} {\bibinfo  {journal} {Phys. Rev. Res.}\ }\textbf {\bibinfo {volume} {6}},\ \bibinfo {pages} {L032059} (\bibinfo {year} {2024})}\BibitemShut {NoStop}%
\bibitem [{\citenamefont {Hughes}\ \emph {et~al.}(2020)\citenamefont {Hughes}, \citenamefont {Sch{\"a}fer}, \citenamefont {Thirumalai}, \citenamefont {Nadlinger}, \citenamefont {Woodrow}, \citenamefont {Lucas},\ and\ \citenamefont {Ballance}}]{hughes2020benchmarking}%
  \BibitemOpen
  \bibfield  {author} {\bibinfo {author} {\bibfnamefont {A.}~\bibnamefont {Hughes}}, \bibinfo {author} {\bibfnamefont {V.}~\bibnamefont {Sch{\"a}fer}}, \bibinfo {author} {\bibfnamefont {K.}~\bibnamefont {Thirumalai}}, \bibinfo {author} {\bibfnamefont {D.}~\bibnamefont {Nadlinger}}, \bibinfo {author} {\bibfnamefont {S.}~\bibnamefont {Woodrow}}, \bibinfo {author} {\bibfnamefont {D.}~\bibnamefont {Lucas}},\ and\ \bibinfo {author} {\bibfnamefont {C.}~\bibnamefont {Ballance}},\ }\href@noop {} {\bibfield  {journal} {\bibinfo  {journal} {Physical Review Letters}\ }\textbf {\bibinfo {volume} {125}},\ \bibinfo {pages} {080504} (\bibinfo {year} {2020})}\BibitemShut {NoStop}%
\bibitem [{\citenamefont {Kj{\ae}rgaard}\ \emph {et~al.}(2000)\citenamefont {Kj{\ae}rgaard}, \citenamefont {Hornekaer}, \citenamefont {Thommesen}, \citenamefont {Videsen},\ and\ \citenamefont {Drewsen}}]{kjaergaard2000isotope}%
  \BibitemOpen
  \bibfield  {author} {\bibinfo {author} {\bibfnamefont {N.}~\bibnamefont {Kj{\ae}rgaard}}, \bibinfo {author} {\bibfnamefont {L.}~\bibnamefont {Hornekaer}}, \bibinfo {author} {\bibfnamefont {A.}~\bibnamefont {Thommesen}}, \bibinfo {author} {\bibfnamefont {Z.}~\bibnamefont {Videsen}},\ and\ \bibinfo {author} {\bibfnamefont {M.}~\bibnamefont {Drewsen}},\ }\href@noop {} {\bibfield  {journal} {\bibinfo  {journal} {Applied Physics B}\ }\textbf {\bibinfo {volume} {71}},\ \bibinfo {pages} {207} (\bibinfo {year} {2000})}\BibitemShut {NoStop}%
\bibitem [{\citenamefont {Lucas}\ \emph {et~al.}(2004)\citenamefont {Lucas}, \citenamefont {Ramos}, \citenamefont {Home}, \citenamefont {McDonnell}, \citenamefont {Nakayama}, \citenamefont {Stacey}, \citenamefont {Webster}, \citenamefont {Stacey},\ and\ \citenamefont {Steane}}]{david_lucas_2004}%
  \BibitemOpen
  \bibfield  {author} {\bibinfo {author} {\bibfnamefont {D.~M.}\ \bibnamefont {Lucas}}, \bibinfo {author} {\bibfnamefont {A.}~\bibnamefont {Ramos}}, \bibinfo {author} {\bibfnamefont {J.~P.}\ \bibnamefont {Home}}, \bibinfo {author} {\bibfnamefont {M.~J.}\ \bibnamefont {McDonnell}}, \bibinfo {author} {\bibfnamefont {S.}~\bibnamefont {Nakayama}}, \bibinfo {author} {\bibfnamefont {J.-P.}\ \bibnamefont {Stacey}}, \bibinfo {author} {\bibfnamefont {S.~C.}\ \bibnamefont {Webster}}, \bibinfo {author} {\bibfnamefont {D.~N.}\ \bibnamefont {Stacey}},\ and\ \bibinfo {author} {\bibfnamefont {A.~M.}\ \bibnamefont {Steane}},\ }\href {https://doi.org/10.1103/PhysRevA.69.012711} {\bibfield  {journal} {\bibinfo  {journal} {Phys. Rev. A}\ }\textbf {\bibinfo {volume} {69}},\ \bibinfo {pages} {012711} (\bibinfo {year} {2004})}\BibitemShut {NoStop}%
\bibitem [{\citenamefont {Leibrandt}\ \emph {et~al.}(2007)\citenamefont {Leibrandt}, \citenamefont {Clark}, \citenamefont {Labaziewicz}, \citenamefont {Antohi}, \citenamefont {Bakr}, \citenamefont {Brown},\ and\ \citenamefont {Chuang}}]{leibrandt_2007}%
  \BibitemOpen
  \bibfield  {author} {\bibinfo {author} {\bibfnamefont {D.~R.}\ \bibnamefont {Leibrandt}}, \bibinfo {author} {\bibfnamefont {R.~J.}\ \bibnamefont {Clark}}, \bibinfo {author} {\bibfnamefont {J.}~\bibnamefont {Labaziewicz}}, \bibinfo {author} {\bibfnamefont {P.}~\bibnamefont {Antohi}}, \bibinfo {author} {\bibfnamefont {W.}~\bibnamefont {Bakr}}, \bibinfo {author} {\bibfnamefont {K.~R.}\ \bibnamefont {Brown}},\ and\ \bibinfo {author} {\bibfnamefont {I.~L.}\ \bibnamefont {Chuang}},\ }\href {https://doi.org/10.1103/PhysRevA.76.055403} {\bibfield  {journal} {\bibinfo  {journal} {Phys. Rev. A}\ }\textbf {\bibinfo {volume} {76}},\ \bibinfo {pages} {055403} (\bibinfo {year} {2007})}\BibitemShut {NoStop}%
\bibitem [{\citenamefont {Hendricks}\ \emph {et~al.}(2007)\citenamefont {Hendricks}, \citenamefont {Grant}, \citenamefont {Herskind}, \citenamefont {Dantan},\ and\ \citenamefont {Drewsen}}]{hendricks2007all}%
  \BibitemOpen
  \bibfield  {author} {\bibinfo {author} {\bibfnamefont {R.}~\bibnamefont {Hendricks}}, \bibinfo {author} {\bibfnamefont {D.}~\bibnamefont {Grant}}, \bibinfo {author} {\bibfnamefont {P.~F.}\ \bibnamefont {Herskind}}, \bibinfo {author} {\bibfnamefont {A.}~\bibnamefont {Dantan}},\ and\ \bibinfo {author} {\bibfnamefont {M.}~\bibnamefont {Drewsen}},\ }\href@noop {} {\bibfield  {journal} {\bibinfo  {journal} {Applied Physics B}\ }\textbf {\bibinfo {volume} {88}},\ \bibinfo {pages} {507} (\bibinfo {year} {2007})}\BibitemShut {NoStop}%
\bibitem [{\citenamefont {Sheridan}\ \emph {et~al.}(2011)\citenamefont {Sheridan}, \citenamefont {Lange},\ and\ \citenamefont {Keller}}]{sheridan2011all}%
  \BibitemOpen
  \bibfield  {author} {\bibinfo {author} {\bibfnamefont {K.}~\bibnamefont {Sheridan}}, \bibinfo {author} {\bibfnamefont {W.}~\bibnamefont {Lange}},\ and\ \bibinfo {author} {\bibfnamefont {M.}~\bibnamefont {Keller}},\ }\href@noop {} {\bibfield  {journal} {\bibinfo  {journal} {Applied Physics B}\ }\textbf {\bibinfo {volume} {104}},\ \bibinfo {pages} {755} (\bibinfo {year} {2011})}\BibitemShut {NoStop}%
\bibitem [{\citenamefont {Vrijsen}\ \emph {et~al.}(2019)\citenamefont {Vrijsen}, \citenamefont {Aikyo}, \citenamefont {Spivey}, \citenamefont {Inlek},\ and\ \citenamefont {Kim}}]{vrijsen_2019}%
  \BibitemOpen
  \bibfield  {author} {\bibinfo {author} {\bibfnamefont {G.}~\bibnamefont {Vrijsen}}, \bibinfo {author} {\bibfnamefont {Y.}~\bibnamefont {Aikyo}}, \bibinfo {author} {\bibfnamefont {R.~F.}\ \bibnamefont {Spivey}}, \bibinfo {author} {\bibfnamefont {I.~V.}\ \bibnamefont {Inlek}},\ and\ \bibinfo {author} {\bibfnamefont {J.}~\bibnamefont {Kim}},\ }\href {https://doi.org/10.1364/OE.27.033907} {\bibfield  {journal} {\bibinfo  {journal} {Opt. Express}\ }\textbf {\bibinfo {volume} {27}},\ \bibinfo {pages} {33907} (\bibinfo {year} {2019})}\BibitemShut {NoStop}%
\bibitem [{\citenamefont {White}\ \emph {et~al.}(2022)\citenamefont {White}, \citenamefont {Low}, \citenamefont {de~Sereville}, \citenamefont {Day}, \citenamefont {Greenberg}, \citenamefont {Rademacher},\ and\ \citenamefont {Senko}}]{brendan_white_2022}%
  \BibitemOpen
  \bibfield  {author} {\bibinfo {author} {\bibfnamefont {B.~M.}\ \bibnamefont {White}}, \bibinfo {author} {\bibfnamefont {P.~J.}\ \bibnamefont {Low}}, \bibinfo {author} {\bibfnamefont {Y.}~\bibnamefont {de~Sereville}}, \bibinfo {author} {\bibfnamefont {M.~L.}\ \bibnamefont {Day}}, \bibinfo {author} {\bibfnamefont {N.}~\bibnamefont {Greenberg}}, \bibinfo {author} {\bibfnamefont {R.}~\bibnamefont {Rademacher}},\ and\ \bibinfo {author} {\bibfnamefont {C.}~\bibnamefont {Senko}},\ }\href {https://doi.org/10.1103/PhysRevA.105.033102} {\bibfield  {journal} {\bibinfo  {journal} {Phys. Rev. A}\ }\textbf {\bibinfo {volume} {105}},\ \bibinfo {pages} {033102} (\bibinfo {year} {2022})}\BibitemShut {NoStop}%
\bibitem [{\citenamefont {Battles}\ \emph {et~al.}(2024)\citenamefont {Battles}, \citenamefont {McMahon},\ and\ \citenamefont {Sawyer}}]{battles_2024}%
  \BibitemOpen
  \bibfield  {author} {\bibinfo {author} {\bibfnamefont {K.~D.}\ \bibnamefont {Battles}}, \bibinfo {author} {\bibfnamefont {B.~J.}\ \bibnamefont {McMahon}},\ and\ \bibinfo {author} {\bibfnamefont {B.~C.}\ \bibnamefont {Sawyer}},\ }\href@noop {} {\bibfield  {journal} {\bibinfo  {journal} {Applied Physics B}\ }\textbf {\bibinfo {volume} {130}},\ \bibinfo {pages} {1} (\bibinfo {year} {2024})}\BibitemShut {NoStop}%
\bibitem [{\citenamefont {Shi}\ \emph {et~al.}(2023)\citenamefont {Shi}, \citenamefont {Todaro}, \citenamefont {Mintzer}, \citenamefont {Bruzewicz}, \citenamefont {Chiaverini},\ and\ \citenamefont {Chuang}}]{shi2023ablation}%
  \BibitemOpen
  \bibfield  {author} {\bibinfo {author} {\bibfnamefont {X.}~\bibnamefont {Shi}}, \bibinfo {author} {\bibfnamefont {S.}~\bibnamefont {Todaro}}, \bibinfo {author} {\bibfnamefont {G.}~\bibnamefont {Mintzer}}, \bibinfo {author} {\bibfnamefont {C.}~\bibnamefont {Bruzewicz}}, \bibinfo {author} {\bibfnamefont {J.}~\bibnamefont {Chiaverini}},\ and\ \bibinfo {author} {\bibfnamefont {I.}~\bibnamefont {Chuang}},\ }\href@noop {} {\bibfield  {journal} {\bibinfo  {journal} {Applied Physics Letters}\ }\textbf {\bibinfo {volume} {122}} (\bibinfo {year} {2023})}\BibitemShut {NoStop}%
\bibitem [{\citenamefont {Kramida}\ \emph {et~al.}(2024)\citenamefont {Kramida}, \citenamefont {{Yu.~Ralchenko}}, \citenamefont {Reader},\ and\ \citenamefont {{and NIST ASD Team}}}]{NIST_ASD}%
  \BibitemOpen
  \bibfield  {author} {\bibinfo {author} {\bibfnamefont {A.}~\bibnamefont {Kramida}}, \bibinfo {author} {\bibnamefont {{Yu.~Ralchenko}}}, \bibinfo {author} {\bibfnamefont {J.}~\bibnamefont {Reader}},\ and\ \bibinfo {author} {\bibnamefont {{and NIST ASD Team}}},\ }\href@noop {} {}\bibinfo {howpublished} {{NIST Atomic Spectra Database (ver. 5.12), [Online]. Available: {\tt{https://physics.nist.gov/asd}} [2025, January 31]. National Institute of Standards and Technology, Gaithersburg, MD.}} (\bibinfo {year} {2024})\BibitemShut {NoStop}%
\bibitem [{\citenamefont {Auchter}\ \emph {et~al.}(2022)\citenamefont {Auchter}, \citenamefont {Axline}, \citenamefont {Decaroli}, \citenamefont {Valentini}, \citenamefont {Purwin}, \citenamefont {Oswald}, \citenamefont {Matt}, \citenamefont {Aschauer}, \citenamefont {Colombe}, \citenamefont {Holz} \emph {et~al.}}]{auchter_2022}%
  \BibitemOpen
  \bibfield  {author} {\bibinfo {author} {\bibfnamefont {S.}~\bibnamefont {Auchter}}, \bibinfo {author} {\bibfnamefont {C.}~\bibnamefont {Axline}}, \bibinfo {author} {\bibfnamefont {C.}~\bibnamefont {Decaroli}}, \bibinfo {author} {\bibfnamefont {M.}~\bibnamefont {Valentini}}, \bibinfo {author} {\bibfnamefont {L.}~\bibnamefont {Purwin}}, \bibinfo {author} {\bibfnamefont {R.}~\bibnamefont {Oswald}}, \bibinfo {author} {\bibfnamefont {R.}~\bibnamefont {Matt}}, \bibinfo {author} {\bibfnamefont {E.}~\bibnamefont {Aschauer}}, \bibinfo {author} {\bibfnamefont {Y.}~\bibnamefont {Colombe}}, \bibinfo {author} {\bibfnamefont {P.}~\bibnamefont {Holz}}, \emph {et~al.},\ }\href@noop {} {\bibfield  {journal} {\bibinfo  {journal} {Quantum Science and Technology}\ }\textbf {\bibinfo {volume} {7}},\ \bibinfo {pages} {035015} (\bibinfo {year} {2022})}\BibitemShut {NoStop}%
\bibitem [{\citenamefont {Weber}\ \emph {et~al.}(2023)\citenamefont {Weber}, \citenamefont {L{\"o}schnauer}, \citenamefont {Wolf}, \citenamefont {Gely}, \citenamefont {Hanley}, \citenamefont {Goodwin}, \citenamefont {Ballance}, \citenamefont {Harty},\ and\ \citenamefont {Lucas}}]{weber_2023}%
  \BibitemOpen
  \bibfield  {author} {\bibinfo {author} {\bibfnamefont {M.}~\bibnamefont {Weber}}, \bibinfo {author} {\bibfnamefont {C.}~\bibnamefont {L{\"o}schnauer}}, \bibinfo {author} {\bibfnamefont {J.}~\bibnamefont {Wolf}}, \bibinfo {author} {\bibfnamefont {M.}~\bibnamefont {Gely}}, \bibinfo {author} {\bibfnamefont {R.}~\bibnamefont {Hanley}}, \bibinfo {author} {\bibfnamefont {J.}~\bibnamefont {Goodwin}}, \bibinfo {author} {\bibfnamefont {C.}~\bibnamefont {Ballance}}, \bibinfo {author} {\bibfnamefont {T.}~\bibnamefont {Harty}},\ and\ \bibinfo {author} {\bibfnamefont {D.}~\bibnamefont {Lucas}},\ }\href@noop {} {\bibfield  {journal} {\bibinfo  {journal} {Quantum Science and Technology}\ }\textbf {\bibinfo {volume} {9}},\ \bibinfo {pages} {015007} (\bibinfo {year} {2023})}\BibitemShut {NoStop}%
\bibitem [{\citenamefont {Lang}\ and\ \citenamefont {Sawyer}(1931)}]{lang1931erste}%
  \BibitemOpen
  \bibfield  {author} {\bibinfo {author} {\bibfnamefont {R.}~\bibnamefont {Lang}}\ and\ \bibinfo {author} {\bibfnamefont {R.}~\bibnamefont {Sawyer}},\ }\href@noop {} {\bibfield  {journal} {\bibinfo  {journal} {Zeitschrift f{\"u}r Physik}\ }\textbf {\bibinfo {volume} {71}},\ \bibinfo {pages} {453} (\bibinfo {year} {1931})}\BibitemShut {NoStop}%
\bibitem [{\citenamefont {Deverall}\ \emph {et~al.}(1953)\citenamefont {Deverall}, \citenamefont {Meissner},\ and\ \citenamefont {Zissis}}]{deverall1953hyperfine}%
  \BibitemOpen
  \bibfield  {author} {\bibinfo {author} {\bibfnamefont {G.}~\bibnamefont {Deverall}}, \bibinfo {author} {\bibfnamefont {K.}~\bibnamefont {Meissner}},\ and\ \bibinfo {author} {\bibfnamefont {G.}~\bibnamefont {Zissis}},\ }\href@noop {} {\bibfield  {journal} {\bibinfo  {journal} {Physical Review}\ }\textbf {\bibinfo {volume} {91}},\ \bibinfo {pages} {297} (\bibinfo {year} {1953})}\BibitemShut {NoStop}%
\bibitem [{\citenamefont {Negnevitsky}\ \emph {et~al.}(2018)\citenamefont {Negnevitsky}, \citenamefont {Marinelli}, \citenamefont {Mehta}, \citenamefont {Lo}, \citenamefont {Fl{\"u}hmann},\ and\ \citenamefont {Home}}]{negnevitsky2018repeated}%
  \BibitemOpen
  \bibfield  {author} {\bibinfo {author} {\bibfnamefont {V.}~\bibnamefont {Negnevitsky}}, \bibinfo {author} {\bibfnamefont {M.}~\bibnamefont {Marinelli}}, \bibinfo {author} {\bibfnamefont {K.~K.}\ \bibnamefont {Mehta}}, \bibinfo {author} {\bibfnamefont {H.-Y.}\ \bibnamefont {Lo}}, \bibinfo {author} {\bibfnamefont {C.}~\bibnamefont {Fl{\"u}hmann}},\ and\ \bibinfo {author} {\bibfnamefont {J.~P.}\ \bibnamefont {Home}},\ }\href@noop {} {\bibfield  {journal} {\bibinfo  {journal} {Nature}\ }\textbf {\bibinfo {volume} {563}},\ \bibinfo {pages} {527} (\bibinfo {year} {2018})}\BibitemShut {NoStop}%
\bibitem [{\citenamefont {Espana}\ \emph {et~al.}(2004)\citenamefont {Espana}, \citenamefont {Joly}, \citenamefont {Hess},\ and\ \citenamefont {Dickinson}}]{Espana2004Caf2}%
  \BibitemOpen
  \bibfield  {author} {\bibinfo {author} {\bibfnamefont {A.~L.}\ \bibnamefont {Espana}}, \bibinfo {author} {\bibfnamefont {A.~G.}\ \bibnamefont {Joly}}, \bibinfo {author} {\bibfnamefont {W.~P.}\ \bibnamefont {Hess}},\ and\ \bibinfo {author} {\bibfnamefont {J.~T.}\ \bibnamefont {Dickinson}},\ }\href {https://www.osti.gov/biblio/15020718} {\bibfield  {journal} {\bibinfo  {journal} {Journal of Undergraduate Research}\ }\textbf {\bibinfo {volume} {4}} (\bibinfo {year} {2004})}\BibitemShut {NoStop}%
\end{thebibliography}%

\end{document}